\documentclass[aps,prd,twocolumn,superscriptaddress,longbibliography,nofootinbib]{revtex4-2}
\usepackage[utf8]{inputenc}
\usepackage{color}
\usepackage{graphicx}
\usepackage{subfigure}
\usepackage{xcolor}
\usepackage[normalem]{ulem}
\usepackage[pdftex,breaklinks,colorlinks,linkcolor=blue,citecolor=teal,anchorcolor=red,urlcolor=cyan]{hyperref}
\usepackage{orcidlink}
\usepackage{multirow}

\usepackage{amsmath,amssymb,amsfonts}

\def\prl{Phys. Rev. Lett.}
\def\prd{Phys. Rev. D}

\def\apj{Astrophys. J.}

\def\apjl{Astrophys. J. Lett.}
\def\aap{Astronomy and Astrophysics}

\def\pau_p{Prog. Theor. Phys.}

\def\mnras{Mon. Not. R. Astron. Soc.}

\def\nat{Nature}

\def\ssr{Space Science Reviews}

\begin{document}
\title{Boosting the growth of intermediate-mass black holes: collisions with massive stars}

\author{Thomas W.~Baumgarte\orcidlink{0000-0002-6316-602X}}
\email{tbaumgar@bowdoin.edu}
\affiliation{Department of Physics and Astronomy, Bowdoin College, Brunswick, Maine 04011, USA}

\author{Stuart L.~Shapiro\orcidlink{0000-0002-3263-7386}}
\email{slshapir@illinois.edu}
\affiliation{Department of Physics, University of Illinois at Urbana-Champaign, Urbana, Illinois 61801}
\affiliation{Department of Astronomy and NCSA, University of Illinois at Urbana-Champaign, Urbana, Illinois 61801}

\begin{abstract}
We perform fully relativistic simulations of the head-on collisions between intermediate-mass black holes and very massive stars.  Such collisions are expected to occur in dense stellar clusters and may play an important role in growing the mass of the seed black hole. For the cases considered here, for which the masses of the black holes and stars are comparable, the vast majority of the stellar material is accreted onto the black hole within a stellar dynamical timescale, as expected from analytical estimates, and leads to a rapid growth of the black hole.  A small amount of mass, which is shock-heated in the wake of the black hole, is ejected from the collision and will contribute to the interstellar material in the cluster.
\end{abstract}

\maketitle

\section{Introduction}

Numerous astronomical observations have firmly established the existence of both stellar-mass black holes, with masses up to around 100 $M_\odot$, and supermassive black holes (SMBHs), with masses of about $10^6 M_\odot$ or more.  

Stellar-mass black holes, which form in the collapse of massive stars at the end of their life cycles, have been observed in X-ray binaries starting with the discovery of Cygnus X-1 (see, e.g., \cite{Oda77} for a discussion and references).  More recently, the direct detection of gravitational waves from merging compact binaries containing black holes has provided a wealth of information about stellar-mass black holes (see, e.g., \cite{GW150914} as well as \cite{ligo} for updates on recent detections).  

SMBHs are believed to exist at the centers of most, if not all galaxies.  From the orbits of stars in its vicinity, the black hole at the center of the Milky Way Galaxy, Sgr A$^*$, has been determined to have a mass of about $4.3 \times 10^6 M_\odot$ (see, e.g., \cite{Schetal02,Gheetal08}).  More recent polarimetric and astrometric observations of flares from Sgr A$^*$ provide consistent values (see \cite{Gravity}), and the Event Horizon Telescope has provided spectacular resolved images of the accretion disk around Sgr A$^*$ (see \cite{EventHorizon}).  SMBHs are also believed to be the central engines of active galactic nuclei (AGNs), which have been observed to high cosmological redshifts (e.g.~\cite{Banetal18}).   In fact, recent observations with the James Webb Space Telescope (JWST) confirm a surprising ubiquity of AGNs at redshifts $z \gtrsim 5$ (see \cite{Greetal24}).  These observations imply that SMBHs must have formed very early in the Universe, challenging our understanding of black-hole growth or formation.

Intermediate-mass black holes (IMBHs), with masses somewhere between those of stellar-mass and SMBHs, may form important stepping stones in the assembly of SMBHs, even though observational evidence for IMBHs is more tenuous than that for stellar-mass or SMBHs.  While some X-ray observations of the irregular galaxy M82 hint at the presence of an IMBH with a mass around $600 M_\odot$ (see, e.g., \cite{Matetal01,Kaaetal01}), later studies suggest that these features could also be explained by a smaller, stellar-mass black hole (e.g.~\cite{Brietal16}).  More promising are recent observations of fast-moving stars in the Milky Way's most massive globular cluster, $\omega$Centauri, which provide evidence for an IMBH with a mass of $8200 M_\odot$ or more (see \cite{Haeetal24}).  

Several different groups have studied possible scenarios for the formation of black holes in stellar clusters (see, e.g., \cite{ShaT85,QuiS87,PorZM02,OmuSH08,DevVRCPZ12,AleN14,GieLHLA15,Map16,ReiSFKB18,AntGG19,Nat21,FraKRS22,KriBS23,VerESR23,FujWTHS24,RanNL24,Deketal24,Krietal24}).  Portegies Zwart and McMillan \cite{PorZM02}, for example, performed $N$-body simulations of clusters and found that repeated collisions between stars may lead to runaway growth, possibly resulting in the formation of an IMBH.  Giersz {\it et.al.}~\cite{GieLHLA15} adopted Monte-Carlo simulations to model the evolution of dense stellar clusters and found that, depending on the number of black holes remaining in the cluster after an early phase of supernova explosions, two different scenarios may lead to the formation of an IMBH.  The authors of \cite{GieLHLA15} also note that the formation of a Bahcall-Wolf cusp \cite{BahW76} depends on several factors, including whether or not the IMBH is located at the cluster's center, and is not always realized in simulations (see also \cite{ZhoBS14}).  Fragione {\it et.al.}~\cite{FraKRS22} used semi-analytical techniques, incorporating full stellar evolution, to demonstrate that some dense stellar clusters have sufficiently deep gravitational potentials to retain many black hole merger remnants despite potentially large recoil kicks acquired in the merger, so that repeated mergers may again result in an IMBH.   Fujii {\it et.al.}~\cite{FujWTHS24} carry out ``star-by-star" simulations of globular clusters in the presence of a parent giant molecular cloud, which deepens the gravitational potential in comparison to cluster simulations in gas-free environments and allows for ongoing star formation.  They find that runaway collisions between stars can lead to the formation of very massive stars (VMSs) with masses exceeding $10^3 M_{\odot}$, which may then collapse to form IMBHs.  The IMBHs formed in such processes may act as the building blocks for the assembly of SMBHs in hierarchical mergers of nuclear clusters (see, e.g., \cite{Krietal24} for a recent discussion), and may trigger a feedback-driven coevolution of SMBHs and galaxies that may help explain recent JWST observations of galaxies at high redshift (see, e.g., \cite{SilBNNW24}).

Mergers and collisions between stars or black holes play a key role in all of the above treatments of cluster evolution, which therefore adopt certain assumptions about the outcome of such interactions.  In particular, it is often assumed that ``100 percent of the mass of a star colliding with a black hole is accreted onto the black hole" (see \cite{GieLHLA15}).  In this paper we explore this assumption using fully relativistic dynamical simulations of head-on collisions between black holes and stars.  Specifically, we consider models of Population III massive stars with masses well above $100 M_\odot$, which we assume to be composed of a hydrogen and helium gas, and we self-consistently evolve both the star and the black hole.  While we find that a small amount of gas that is shock-heated in the wake of the black hole is ejected from the star, the vast majority of the stellar material is indeed accreted onto the black hole in all the cases that we studied, confirming the above assumption.

Massive stars and black holes are characterized by vastly different time and length scales, which makes self-consistent dynamical simulations of their collisions very demanding computationally. In order to make such simulations feasible with our computational resources we adopt relatively compact models of massive stars.  We perform simulations for different stellar compactions and mass ratios and find that, while our results do show some dependence on these parameters quantitatively, they agree qualitatively in that the vast majority of the stellar matter is accreted by the black hole during the black hole's first transit through the star.  We nevertheless caution that our simulations, which we believe are the first of this kind in full general relativity, cover only a small part of the parameter space.  Our results for head-on collisions should also be distinguished from tidal and grazing encounters, in which much more mass may be ejected, and much less mass accreted by the black hole (see, e.g., \cite{MetS16} and references therein).  

Our paper is organized as follows.  We start with analytic estimates in Section \ref{sec:estimates}, which provide insight into the dynamics and likely outcome of collisions between black holes and stars.  In Section \ref{sec:numerics} we discuss details of our numerical approach, including our treatment of the stellar equation of state, the construction of initial data, numerical methods for the dynamical evolution, and diagnostics.  We describe our numerical results in Section \ref{sec:results}, and conclude with a discussion in Section \ref{sec:summary}.  We also include three appendices with details of some of our calculations.  We use cgs units in most of this paper, but adopt geometrized units with $G = 1 = c$ in Section \ref{sec:diagnostics} and Appendix \ref{app:EOS_num}.

\section{Analytic Estimates}
\label{sec:estimates}

In this section we anticipate the results of our numerical simulations by providing a few rough Newtonian analytic estimates.  

Upon impact, the BH enters the VMS with an initial relative speed 
\begin{equation} \label{v0}
v_0 \simeq v_{\rm esc} = \left( \frac{2 G M}{R_*} \right)^{1/2},
\end{equation}
where the total mass $M = m + M_*$ is the sum of the black-hole mass $m$ and the stellar mass $M_*$, $R_*$ is the stellar radius, and where we have assumed that we can neglect the relative speed at large separation, $v_\infty \ll v_{\rm esc}$.  Once inside the star, kinetic energy is dissipated by both hydrodynamical friction and accretion drag.  In the supersonic regime the former is likely to dominate (see Appendix \ref{app:fric_and_drag} for a justification) and can be approximated as
\begin{equation} \label{F_defl}
    F_{\rm defl} \simeq \frac{4 \pi \rho G^2 m^2}{v^2}  \ln \Lambda,
\end{equation}
(see, e.g., \cite{Cha43,RudS71}; see also \cite{Ost99} for extension to the subsonic regime). As we show in Appendix \ref{app:mach}, the black hole would remain supersonic throughout the star in the absence of frictional forces.  In (\ref{F_defl}) $\rho$ is the density and $\ln \Lambda$ is the Coulomb logarithm, which, following several previous authors, we will approximate as $\ln \Lambda \simeq 10$ in the following (e.g.~\cite{Abretal09}).  

\begin{figure}
    \centering
    \includegraphics[width = 0.49 \textwidth]{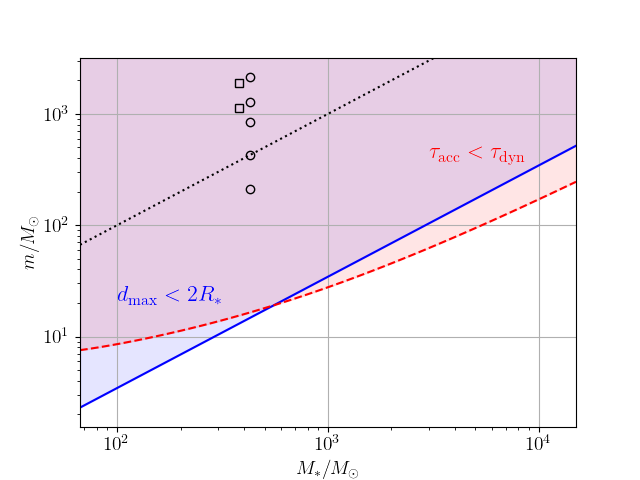}
    \caption{Estimates regarding the result of a direct collision between and IMBH of mass $m$ with a VMS of mass $M_*$.  For combinations of the two masses above the solid (blue) line the BH wile not re-emerge from the star [see Eq.~(\ref{d_max})], and for those above the dashed (red) line the entire star will be accreted by the BH on timescales shorter than the stellar dynamical timescale [see Eq.~(\ref{taus})].  The dotted (black) line marks equal masses $m = M_*$, and the open squares and circles mark configurations for which we performed dynamical simulations.}
    \label{fig:estimate}
\end{figure}

If we were to assume that the black hole loses only very little of its initial kinetic energy relative to the star, whereby $K_0 = \mu v_0^2/2 \simeq G m M_* / R_*$ (where $\mu = m M_* / M$ is the reduced mass) during its passage through the star, we could estimate $F_{\rm defl}$ by setting $v \simeq v_0$ and $\rho \simeq \bar \rho = 3 M_* / (4 \pi R_*^3)$ in Eq.~(\ref{F_defl}).  The total energy lost, after having traversed the star and traveled a distance $d = 2 R_*$, would then be
\begin{equation}
\Delta E \simeq F_{\rm defl} d \simeq \frac{6 G M_* m^2}{R_*} \ln \Lambda \simeq 6 \ln \Lambda \frac{m}{M_*} K_0.
\end{equation}
For $m \simeq M_*$ we have $\Delta E \gg K_0$, clearly contradicting our assumption above.  Instead, we can crudely estimate the maximum distance traveled by equating $\Delta E$ with $K_0$, which yields
\begin{equation} \label{d_max}
\frac{d_{\rm max}}{2 R_*}
\lesssim \frac{1}{3 \ln \Lambda} \frac{M_*}{m}.
\end{equation}

For sufficiently large $m$ we have $d_{\rm max} < 2 R_*$, so that the BH is unlikely to reemerge from the SMS, and will instead keep accreting stellar material as an ``endoparasitic" BH inside a host star.  In Fig.~\ref{fig:estimate} the blue solid line marks $d_{\rm max} = 2 R_*$; for BH masses above this line the BH will not reemerge from the SMS.

In this paper we are mostly concerned with BHs
with masses comparable to stellar masses, $m \simeq M_*$, since it was the collapse of the stars that presumably gave birth to the black holes in the first place. However, we note that even when the BH mass is appreciably less than the stellar mass and the BH emerges from the star, the BH will still be bound
to the star, will reenter the star on successive orbits and eventually
be captured, and ultimately consume the star.

For BHs with mass $m \simeq M_*$ our estimates above suggest that the BH will be absorbed inside the SMS during the initial encounter.  The process of absorbing the BH adds two new contributions to the SMS's energy budget: the kinetic energy $K_0 \simeq \mu v_0^2/2$, much of which has been converted into heat by dynamical friction, and a gravitational interaction energy $E_{\rm int} \simeq - G m M_* / R_*$ between the BH and its host.   Under the assumption that $v_\infty \ll v_{\rm esc}$ the two contributions nearly cancel each other, so that the stellar binding energy remains negative.  Accordingly, it appears unlikely that the collision could lead to an explosion and complete disruption of the star, although the stellar mass likely will be redistributed, leading to a few puffs of
outgoing gas together with some concentration of mass near the BH. Hence, our crude estimates suggest that most of the stellar material will remain bound to, and will ultimately be accreted by, the black hole.

The above arguments leave two possible scenarios: either the entire star collapses into the black hole promptly upon the collision, or the collision leads to the formation of a transient, quasi-equilibrium ``Thorne-\.Zytkow-like" object, consisting of a SMS hosting a BH in its core.  Such an object is certainly unstable on secular accretion timescales, but might be stable on dynamical timescales.  In order to distinguish the two scenarios we compare the mean stellar dynamical time scale 
\begin{equation} \label{tau_dyn}
\tau_{\rm dyn} \simeq \left( \frac{R_*^3}{G (M_* + m)} \right)^{1/2}
\end{equation}
with the accretion timescale
\begin{equation} \label{tau_acc}
\tau_{\rm acc} \simeq \frac{m}{\dot m} 
= \frac{a_s^3}{4 \pi \lambda G^2 m \rho}.
\end{equation}
Here we have assumed that the accretion rate $\dot m$ can be estimated from spherical, steady-state Bondi accretion
\begin{equation} \label{Bondi}
\dot m \simeq \dot m_{\rm sph} = \frac{4 \pi \lambda G^2 m^2 \rho}{a_s^3}
\end{equation}
(see \cite{Bon52}; see also \cite{ShaT83} for a textbook treatment), and that the relative speed between the BH and SMS has dropped well below the sound speed, in accordance with our estimates above.  Below we will further assume that the BH has come to rest close the the center of the star.  In Eqs.~(\ref{tau_acc}) and (\ref{Bondi}), $a_s$ is the sound speed and $\lambda$ the accretion eigenvalue, which, for adiabatic indices $\Gamma \leq 5/3$, is given by
\begin{equation} \label{lambda}
    \lambda = \frac{1}{4} \left( \frac{2}{5 - 3 \Gamma} \right)^{(5 - 3 \Gamma)/(2(\Gamma - 1))}
\end{equation}
(see \cite{ShaT83}).  At the center of the star we may approximate $a_s \simeq (\Gamma G M_*/R_*)^{1/2}$ and $\rho = \rho_c = \delta \bar \rho$, where $\delta$ is the central condensation.  Combining the above we find
\begin{equation} \label{taus}
\frac{\tau_{\rm acc}}{\tau_{\rm dyn}} 
\simeq \frac{\Gamma^{3/2}}{3 \lambda \delta} \frac{M_*}{m} \left(1 + \frac{m}{M_*} \right)^{1/2}.
\end{equation}

In order to evaluate Eq.~(\ref{taus}) we first observe that in massive stars, which are dominated by radiation pressure $P_r$ and in which the gas pressure $P_g$ plays the role of a small perturbation, we may estimate $\Gamma$ by 
\begin{equation} \label{Gamma}
    \Gamma \simeq \frac{4}{3} + \frac{\beta}{6},
\end{equation}
where 
\begin{equation} \label{beta}
    \beta \equiv \frac{8 k_B}{s} \simeq \frac{P_g}{P_r} \simeq 8.46 \left( \frac{M_*}{M_\odot} \right)^{-1/2}
\end{equation}
(see, e.g., Chapter 17 in \cite{ShaT83}).  In (\ref{beta}), $k_B$ is the Boltzmann constant and $s$ the entropy per baryon.   For a given value of $M_*$ we estimate $\Gamma$ from (\ref{Gamma}), then get the accretion eigenvalue $\lambda$ from (\ref{lambda}), and finally find $\delta$ by interpolating between tabulated data (see Table \ref{tab:lane_emden} in Appendix \ref{app:mach}).  

In Fig.~\ref{fig:estimate}, the red dashed line marks the black hole masses $m$ for which $\tau_{\rm acc} \simeq \tau_{\rm dyn}$; for $m$ above this line, the accretion timescale is shorter than the dynamical timescale, suggesting a prompt collapse rather than the formation of a transient, dynamically stable object. In order to find this line, we set $\tau_{\rm acc} = \tau_{\rm dyn}$ in Eq.~(\ref{taus}) and write the remaining equation as a quadratic equation for the mass ratio $q \equiv m/M_*$.  The positive solution for $q$ can then be written as 
\begin{equation}
q = \eta \left(1 + \frac{\eta^2}{4}\right)^{1/2} + \frac{\eta^2}{2},
\end{equation}
where we have abbreviated $\eta = \Gamma^{3/2}/(3 \lambda \delta)$.

In summary, our estimates suggest that the direct collision of a IMBH with a VMS of comparable mass will lead to a prompt and rapid accretion of nearly the entire star into the IMBH, thereby boosting the growth of the BH's mass.  In the following we confirm these expectations with fully dynamical simulations in general relativity.

\section{Numerical Setup}
\label{sec:numerics}

\subsection{The equation of state}
\label{sec:eos}

We assume that the equation of state (EOS) governing the VMSs considered in this paper is dominated by contributions from 
blackbody radiation and Maxwell-Boltzmann gas.  Assuming that photons only contribute to the former we write the radiation pressure as
\begin{equation} \label{P_r}
P_r = \frac{1}{3} a T^4,
\end{equation}  
where $T$ the temperature and $a$ the radiation constant
\begin{equation}
a = \frac{8 \pi^5 k_B^4}{15 c^3 h^3}.
\end{equation} 
Here $k_B$ is the Boltzmann constant and $h$ is Planck's constant.
The gas pressure is given by 
\begin{equation} \label{P_g}
P_g = \frac{\rho_0 k_B T}{\mu m_B},
\end{equation}
where $\rho_0$ is the rest-mass density, $m_B$ the baryon mass, and $\mu$ the mean molecular weight.  We assume that the gas has negligible metallicity, and hence consider contributions from hydrogen and helium only. Denoting the mass fraction of hydrogen by $X$ and that of helium by $Y = 1 - X$, and further assuming that the gas is fully ionized (which is consistent with the temperatures in the stellar models adopted below) we have
\begin{equation} \label{mu}
\frac{1}{\mu} = 2X + \frac{3Y}{4}.
\end{equation}
The total pressure is then given by
\begin{equation} \label{pressure_1}
P = P_r + P_g = \frac{1}{3} a T^4 + \frac{\rho_0 k_B T}{\mu m_B}.
\end{equation}

Given the above assumptions, the specific energy density $\epsilon$ is
\begin{equation} \label{epsilon_1}
\epsilon = \epsilon_r + \epsilon_g = \frac{a T^4}{\rho_0} + \frac{3}{2}\frac{k_B T}{\mu m_B}.
\end{equation}
The internal energy density $\rho_i$ is then given by $\rho_i = \epsilon \rho_0$, and the total mass-energy density is
\begin{align} \label{density_1}
\rho = \rho_0 ( 1 + \epsilon)
= \rho_0 + a T^4 + \frac{3}{2} \frac{\rho_0 k_B T}{\mu m_B}.
\end{align}

Finally, the total entropy per baryon is given by
\begin{align} \label{entropy_1}
s & = s_r + s_g 
= \frac{4 a m_B T^3}{3 \rho_0}
+ \frac{k_B}{\mu} \ln \frac{(k_B T)^{3/2}}{\rho_0} + s_0
\end{align}
where $s_0$ is a constant.  We provide a value of this constant, together with a sketch of its derivation, in Appendix \ref{sec:EOS_num:entropy} (see Eq.~(\ref{app_entropy_constant})).

In the limiting case that contributions from radiation dominate over those from gas we may write 
\begin{equation}
P \simeq P_r = \frac{1}{3} a T^4
\simeq \frac{a}{3} \left(\frac{3 s_r \rho_0}{4 a m_B} \right)^{4/3} \simeq K \rho_0^{4/3},
\end{equation}
where we have defined 
\begin{equation} \label{K_1}
K \equiv \frac{a}{3} \left(\frac{3 s}{4 a m_B} \right)^{4/3}
\end{equation}
and assumed $s_r \simeq s$.  Further assuming that the star is isentropic initially, with $s = s_{\rm init} = const$, we have $K = const$ as well, so that the stellar structure approaches that of an $n=3$ ($\Gamma = 4/3$) polytrope in this limit.

We do not take into account nuclear reactions in our simulations.  The stellar models that we adopt in Section \ref{sec:init:OV} below feature temperatures that are high enough to ignite nuclear reactions, so that a proper treatment of nuclear reactions would likely have some effect on the overall structure of the star.  However, the reactions are very unlikely to have a dynamical effect on the short timescales considered here, so that they should not affect our findings qualitatively.  We similarly ignore pair production, which would also be triggered in the specific models that we adopt below -- just marginally in our initial data, but more so during the evolution in the shock-heated wake of the black hole.  Pair production involves the conversion of internal thermal energy into the rest mass-energy of electrons and positrons, and the corresponding decrease in the thermal pressure is expected to make collapse more likely.   Since we find that almost the entire star is accreted by the BH in the collisions that we consider in Sect.~\ref{sec:results} even in the absence of pair production, including the latter is likely to further strengthen our conclusion.

We provide details on the numerical evaluation of the EOS in Appendix \ref{app:EOS_num}. We also note that we solve Eq.~(\ref{epsilon_1}) analytically, which, for given values of $\epsilon$ and $\rho_0$, is a quartic equation for the temperature $T$, as discussed in \cite{BauS25a}. 

\subsection{Initial models}
\label{sec:initialmodels}

\subsubsection{Oppenheimer-Volkoff models}
\label{sec:init:OV}

We start by constructing spherically symmetric equilibrium solutions for stars supported by the EOS of Sect.~\ref{sec:eos}.  Specifically, we first choose a certain {\em target mass} $M_{\rm tar}$ for the sole purpose of providing an estimate for the initial entropy per baryon $s_{\rm init}$ from (\ref{beta}),
\begin{equation} \label{s_init}
s_{\rm init} = \frac{8 k_B}{\beta} = \frac{8 k_B}{8.46} \left( \frac{M_{\rm tar}}{M_\odot} \right)^{1/2}.
\end{equation}
We then solve the Oppenheimer-Volkoff (OV) equations \cite{OppV39} to construct isentropic stellar models.  These equations include a differential equation for the pressure.  In practice we choose a central rest-mass density $\rho_{0c}$ and set $s = s_{\rm init}$ to solve (\ref{entropy_1}) for the central temperature, which we then insert into (\ref{pressure_1}) to find the central pressure.  Using these as starting values we integrate the OV equations to get the pressure at fixed entropy $s = s_{\rm init}$.  For a given pressure and entropy we invert Eqs.~(\ref{pressure_1}) and (\ref{entropy_1}) at each point to get the rest-mass density $\rho_0$ and the temperature $T$.  From these, we can find $\rho$ using (\ref{density_1}) and thereby continue the integration.

\begin{figure}[t]
    \centering
    \includegraphics[width=0.45 \textwidth]{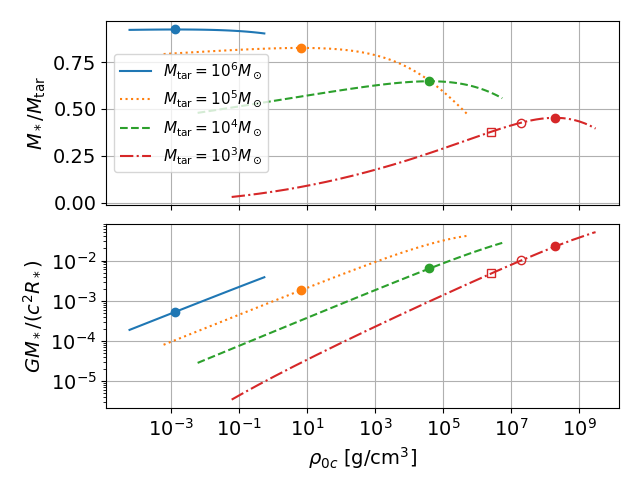}
    \caption{The gravitational mass $M_*$ (in units of the target mass $M_{\rm tar}$, upper panel) and the compaction $GM_*/(c^2 R_*)$ (lower panel) as a function of central rest-mass density $\rho_{0c}$ for isentropic massive and supermassive stars governed by a mixture of gas and radiation pressure.  The different lines correspond to curves of different constant entropy $s_{\rm init}$, specified by the  differernt target masses $M_{\rm tar}$ [see Eq.~(\ref{s_init})].  The solid dots denote the maximum-mass configuration for a given value of $M_{\rm tar}$; the open circles and squares mark the stable VMS models  A and B adopted in our dynamical simulations (see also Table \ref{tab:star}).}
    \label{fig:OV_models}
\end{figure}

For a given $M_{\rm tar}$, and hence $s_{\rm init}$, we construct stellar equilibrium models for a range of central rest-mass densities $\rho_{0c}$.  For each value of $\rho_{0c}$ we obtain the stellar (gravitational) mass $M_*$ together with the stellar (areal) radius $R_*$ for the resulting equilibrium configuration.  In Fig.~\ref{fig:OV_models} we show the masses $M_*$ as well as the compactions $GM_*/(c^2 R_*)$ as a function of $\rho_{0c}$ for different values of $M_{\rm tar}$. Along each curve of constant entropy $s = s_{\rm init}$, labeled by $M_{\rm tar}$, we mark the maximum-mass configuration with a solid dot; these dots also separate stable configurations (for lower densities) from unstable configurations (for higher densities).  We note that for each value of $M_{\rm tar}$ the mass of the maximum-mass configuration is somewhat smaller than $M_{\rm tar}$, but that we recover $M_* \rightarrow M_{\rm tar}$ in the limit $M_{\rm tar} \rightarrow \infty$, in which case $P_g / P_r \rightarrow 0$ and our previous estimates for the entropy become accurate.


Fully relativistic dynamical simulations of these stars are challenging for several reasons.  We first note that the ratio between the dynamical timescale (\ref{tau_dyn}) (with $m = 0$) and the light-crossing time $\tau_{\rm lc} \simeq R_* / c$ scales with the inverse square root of the compaction,
\begin{equation}
\frac{\tau_{\rm dyn}}{\tau_{\rm lc}} \simeq \left( \frac{GM_*}{c^2 R_*} \right)^{-1/2},
\end{equation}
so that typical simulations of low-compaction stars require many more timesteps than those of compact stars.  Moreover, massive stars are well approximated by $\Gamma = 4/3$ polytropes and therefore have small binding energies $M_0 - M_*$, where $M_0$ is the stellar rest mass.  As a consequence, even a small numerical error can induce relatively large changes in the star.  Finally, these stars are centrally condensed, so a high numerical resolution around the stellar center is required to reduce such numerical errors.  The last point is particularly relevant for our numerical setup, in which we adopted spherical polar coordinates centered on the black hole (see Sect.~\ref{sec:dynamics} below).

\begin{table}[t]
    \centering
    \begin{tabular}{c|c|c}
        & Star A & Star B \\
        \hline
        $M_*~[M_\odot]$ & $427$  & 378 \\
        $M_0~[M_\odot]$ & $430$  & 380 \\
        $R_*~[\mbox{cm}]$ & $6.35 \times 10^9$     & $1.18 \times 10^{10}$ \\
        $G M_*/(c^2 R_*)$ &  $0.01$ & $0.0048$ \\
        $\rho_{0c}~[\mbox{g cm}^{-3}]$  & $1.97 \times 10^{7}$  & $2.60 \times 10^{6}$ \\ 
        $T_c~[\mbox{K}]$ & $1.37 \times 10^{10}$ & $6.76 \times 10^9$ \\
        $\tau_{\rm dyn}~[\mbox{s}]$ & 2.1 & 5.7
        \end{tabular}
    \caption{Physical parameters of the fiducial stellar models adopted in the initial data for our simulations, namely the stellar (gravitational) mass $M_*$, the stellar rest mass $M_0$, the stellar (areal) radius $R_*$, the compaction $G M_*/(c^2 R_*)$, the central rest-mass density $\rho_{0c}$, and the central temperature $T_c$.  As a rough estimate of the stellar dynamical timescale we evaluate (\ref{tau_dyn}) with $m = 0$ in the last row.  Star A is marked by the open circles in Figs.~\ref{fig:estimate} and \ref{fig:OV_models}, and star B by the open squares.}
    \label{tab:star}
\end{table}

In this paper we adopt fiducial stellar models with $M_{\rm tar} = 10^3 M_\odot$, for which the stable branch in the top panel of Fig.~\ref{fig:OV_models} extends to relatively large values of the compactions.  Further noticing that the estimates of Section \ref{sec:estimates} are independent of the stellar radius, so that the choice of the latter should not affect the outcome qualitatively, we choose two models relative close to the maximum-mass configuration.  We mark these fiducial models with the open squares and circles in Fig.~\ref{fig:OV_models} and list their physical properties in Table~\ref{tab:star}.   While these models, with their high densities and temperatures and small radii, may not be realistic models for VMSs (see also the discussion in Section \ref{sec:eos}), more realistic models would further increase the range in both time and length scales, and would make it even harder to simulate collisions with black holes.   Given the constraints on our numerical resources we therefore content ourselves with these admittedly extreme models of VMSs in our initial foray into performing fully relativistic and self-consistent simulations of their collisions with black holes.

\subsubsection{Initial data for head-on collisions}
\label{sec:init:headon}

We adopt the approach discussed in \cite{BauS24d} to construct initial data that describe the head-on collision of a BH with a VMS.  Specifically, we solve the equations in a frame in which the star is at rest momentarily, so that, assuming conformal flatness and maximal slicing, the momentum constraint can be solved analytically using a Bowen-York solution \cite{BowY80} to describe the black hole with given momentum ${\mathcal P}$.  We then adopt the puncture method of \cite{BraB97} and write the conformal factor as
\begin{equation}
\psi = \psi_{*} + \psi_{\rm BH} + u - 1
= \psi_{*} + \frac{\mathcal M}{r} + u.
\end{equation}
Here $\psi_{*}$ describes the stellar contribution to the conformal factor and is found from the Oppenheimer-Volkoff solution of Sect.~\ref{sec:init:OV}, ${\mathcal M}$ is the black hole's puncture mass, and $r$ measures the (coordinate) distance from the black hole's center, which is placed a certain distance $l$ from the stellar center.  We then solve the Hamiltonian constraint for the correction $u$, which completes a solution to Einstein's constraint equations.   

For a given value of ${\mathcal P}$ we iterate over the star's central rest-mass density $\rho_{0c}$ and the puncture mass ${\mathcal M}$ until the black hole's horizon mass $m$ [see Eq.~(\ref{m_irr}) below] and the stellar rest mass $M_0$ take the desired values.  Finally, we choose ${\mathcal P}$ so that it approximates the (relative) free-fall speed from rest at infinity at distance $l$ from the stellar center. As expected for the weakly relativistic stars considered here, a Newtonian estimate yields values very similar to the relativistic procedure described in \cite{BauS24d} and applied here. 

\begin{table*}[]
    \centering
    \begin{tabular}{c|c|c|c|c|c|c|c||c|c}
        Stellar model & $M_*~[M_\odot]$ & $GM_*/(c^2 R_*)$ & Simulation & $m(0)~[M_\odot]$ & $m(0)/M_0(0)$ & $l/R_*$ & ${\mathcal P}/(m(0)c)$ & $m_{\rm fin}~[M_\odot]$ & $M_0^{\rm ext} / M_0(0)$ \\
        \hline
         \multirow{5}{*}{A} & \multirow{5}{*}{427} & \multirow{5}{*}{0.01}
           & A1 & 215 & 0.5 & 1.18 & 0.164  & 636    & $<4.1\%$ \\
         & & & A2 & 430 & 1.0 & 1.20 & 0.186 & 854 & $<1.6\%$ \\
         & & & A3 & 860 & 2.0 & 1.91 & 0.181 & 1290 & $<0.6\%$\\
         & & & A4 & 1290 & 3.0 & 2.38 & 0.189 & 1714 & $<0.3\%$\\        
         & & & A5 & 2150 & 5.0 & 2.84 & 0.213 & 2571 & $<0.004\%$\\
         \hline
         \multirow{2}{*}{B} & \multirow{2}{*}{378} & \multirow{2}{*}{0.0048}
           & B1 & 1140 & 3.0 & 1.27 & 0.178 & 1512 & $<1.9\%$ \\
         & & & B2 & 1900 & 5.0 & 1.52 & 0.202 & 2277 & $<1.3\%$ \\ 
    \end{tabular}
    \caption{Summary of our numerical simulations.  For each of the two stellar models A and B (see also Table \ref{tab:star}) we simulate collisions with black holes of different initial mass $m(0)$.  We specify the initial conditions by choosing the initial separation $l$ from the stellar center together with the initial momentum ${\mathcal P}$ (see text for details).  In the two right-most columns we list the black hole mass $m_{\rm fin}$ at the end of our simulations, as well as $M_0^{\rm ext} / M_{0}(0)$, i.e.~the fraction of the stellar rest mass that remains outside the black hole horizon.  Animations of selected simulations can be found at \cite{animation}. }
    \label{tab:indata}
\end{table*}

We note that the above procedure does not account for the tidal deformation of the star.  In \cite{BauS24d}, where we considered black holes with masses significantly smaller than those of the star, we were able to neglect these tidal deformations and constructed initial data with the black hole at the stellar surface.  Here, however, the two masses are similar, and we therefore start the simulations with the black hole a certain distance away from the star, which allows the star to develop a tidal bulge before the impact without making the simulation runtimes prohibitively long.  

We list parameters for our initial data models in Table~\ref{tab:indata}, and show examples of initial density and temperature profiles in the upper left panels of Figs.~\ref{fig:density} and \ref{fig:temperature}.

\subsection{Dynamical Evolution}
\label{sec:dynamics}

We perform dynamical evolution simulations with the numerical code originally described in \cite{BauMCM13}.  Specifically, the code implements the Baumgarte-Shapiro-Shibata-Nakamura (BSSN) formulation \cite{NakOK87,ShiN95,BauS99} of Einstein's equations in spherical polar coordinates.  All singular terms at the origin (where $r = 0$) and on the axis (where $\sin \theta = 0$) are handled analytically with the help of a reference-metric approach (see, e.g., \cite{BonGGN04,ShiUF04,Bro09,Gou12}) as well as a rescaling of all tensorial quantities.  We apply similar techniques to the equations of relativistic hydrodynamics \cite{MonBM14,BauMM15}, and use a Harten-Lax-van-Leer-Einfeldt approximate Riemann solver \cite{HarLL83,Ein88} with a simple monotonized central-difference limiter reconstruction scheme \cite{Van77}. We treat the vacuum in the stellar exterior by using an artificial atmosphere with $\rho_0^{\rm atm} = 10^{-8} \rho_0^{\rm max}(0)$ there, and excise the evolution of the matter (but not that of the gravitational fields) on the innermost few gridpoints around the black hole's center, well inside the black hole horizon. For the simulations presented here we use fourth-order finite-differencing for spatial derivatives, together with a fourth-order Runge-Kutta method for the time evolution.  

We place the black hole at the origin and keep it there throughout the evolution (using the coordinate conditions discussed below), while the star's center starts out on the axis towards the south pole ($\theta = \pi$; see Figs.~\ref{fig:density} and \ref{fig:temperature} for examples).  Although our code does not assume any symmetries, we can take advantage of axisymmetry in these head-on collisions by using only one grid point for the azimuthal angle $\varphi$ (plus ghost zones surrounding it).   In order to achieve adequate resolution of both the black hole and the much larger star, we use non-uniform grids for both the radius $r$ and the polar angle $\theta$ as described below.  

Our radial grid extends from the origin at $r = 0$ to an outer boundary at $r = r_{\rm out}$ and is constructed from a uniform, cell-centered grid with $N_r$ grid points in an auxiliary variable $0 \leq x \leq 1$ with the map
\begin{equation}
r = r_{\rm out} \frac{\sinh(s_r x)}{\sinh(s_r)},
\end{equation}
where $s_r$ is a constant parameter that governs the ``non-uniformity" of the grid (see \cite{RucEB18}).  For $s_r$ we recover a uniform grid in $r$, while for $s_r > 0$ the grid is nearly uniform close to the origin but becomes approximately logarithmic at large distances. For the results reported here we use $N_r = 384$ grid points, place the outer boundary at $r_{\rm out} = 1.49\times 10^{11}$ cm, and adjusted $s_r$ according to the initial black-hole mass so that the black-hole interior is always covered by at least about 12 grid points.

The (polar) angular grid extends from $\theta = 0$ to $\theta = \pi$, corresponding to the two poles. We construct this grid from a uniform, cell-centered grid with $N_\theta$ grid points in an auxiliary variable $0 \leq y \leq 1$ using the map
\begin{equation}
\theta = \pi y + s_\theta \sin( 2 \pi y),
\end{equation}
where $s_\theta$ is again a constant parameter.  For $s_\theta = 0$ the grid is uniform, while for negative $s_\theta$ the grid resolution is finer in the vicinity of the poles than at the equator.  Since our simulations require a higher grid resolution towards the poles than towards the equator (to better resolve both the stellar center at $\theta = \pi$ and the black hole's wake around $\theta = 0$) we use $s_\theta = -0.32$ together with $N_\theta = 64$ for the simulations shown here.

We impose coordinate conditions by evolving the lapse function $\alpha$ with the 1+log slicing condition 
(see \cite{BonMSS95}), starting with a pre-collapsed lapse $\alpha = \psi^{-2}$ at the initial time, and the  the shift vector $\beta^i$ using a modified Gamma-driver condition \cite{AlcBDKPST03}, starting with $\beta^i = 0$ initially.  As described in detail in \cite{BauS24d} (see Section II.B.2), we modify the Gamma-driver condition by adding, at each time step, a vector that (i) is approximately constant in the vicinity of the black hole, and (ii) results in the shift remaining zero at the center of the black hole.  Because of the first condition, the new gauge condition inherits all the desirable properties of the Gamma-driver condition, while the second condition ensures that the black hole itself will remain at its original coordinate location.  This latter property is crucial for our simulations, because it allows us to keep the black hole at the origin of the coordinate system and take advantage of the high grid resolution there throughout the evolution. 

\subsection{Diagnostics}
\label{sec:diagnostics}

We locate apparent horizons in both our initial data and the evolved spacetimes using the method of \cite{Shi97,ShiU00}.  We then compute the proper area $\mathcal{A}$ of the horizon (see, e.g., \cite{BauCSST96} as well as Appendix C in \cite{BauS10}), and from it an approximate value of the black hole's irreducible mass
\begin{equation} \label{m_irr}
 m = \left( \frac{\mathcal{A}}{16 \pi} \right)^{1/2},
\end{equation}
where we now adopt geometrized units with $G = 1 = c$.  In the absence of black hole spin, which is the case
here, the irreducible mass is also the isolated horizon mass (see, e.g., \cite{AshK03}).  We next compute the accretion rate 
\begin{equation} \label{accretion_rate}
\dot m_0 = - \oint_{\mathcal{H}} \alpha \rho_0 u^i dS_i
\end{equation}
from the flux of rest-mass across the black hole horizon $\mathcal{H}$.  Here $u^i$ is the fluid four-velocity and $dS_i$ the outward-oriented surface element on $\mathcal{H}$ (see Section II.C in \cite{BauS24d} for details).  We note that the accretion rate (\ref{accretion_rate}) measures the rate at which rest mass enters the black hole, which is not the same as the rate at which the black hole's total mass-energy (\ref{m_irr}) increases (see \cite{AguST21,AguTSL21} as well as the Appendix in \cite{BauS24d} for discussion).

During the dynamical evolution we track the total rest mass remaining outside the black hole horizon from
\begin{equation} \label{M_ext}
M_0^{\rm ext} = \int_{\rm ext} d^3 x \sqrt{\gamma} \alpha u^t \rho_0,
\end{equation}
where the integral is taken over the black hole exterior only.  In (\ref{M_ext}) $\gamma$ is the determinant of the spatial metric and $u^t$ the contravariant time component of the fluid four-velocity (see, e.g., Section 3.5 in \cite{BauS10}).  

We also provide a crude estimate for the amount of mass that is unbound by the black hole.  A particle that interacts only gravitationally follows a geodesic.  In stationary spacetimes the covariant time component $u_t$ of such a particle's four-velocity is a constant of motion and defines the particle's conserved total energy $E = - \mu u_t$, where $\mu$ is the particle's mass.  For the particle to be able to escape to infinity, its energy must exceed its rest mass $\mu$; therefore, an unbound particle must have $u_t < -1$.  Late in our evolutions the spacetime at $r \gtrsim R$ approaches stationarity, and in low-density regions the forces on the fluid are dominated by gravitational forces.  We may therefore estimate the amount of material $M_0^{\rm unbound}$ that is unlikely to be accreted by the black hole by restricting the integral (\ref{M_ext}) to fluid particles that have $u_t < - 1$ and move away from the black hole, $u^r > 0$, in addition to being outside the black hole.

\section{Results}
\label{sec:results}

\begin{figure*}
    \centering
    \includegraphics[width=0.36\textwidth]{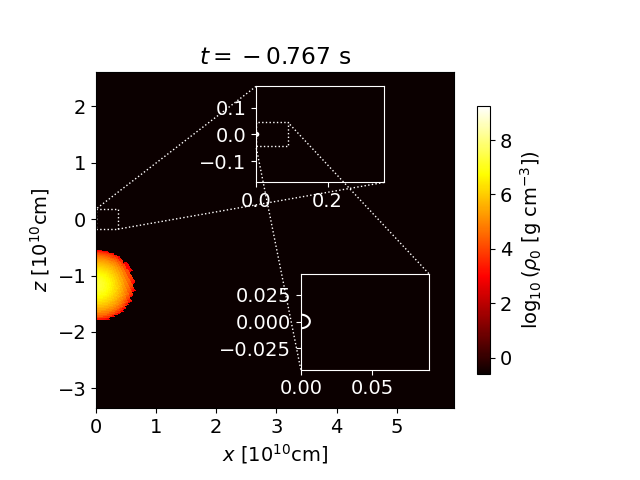}
    \includegraphics[width=0.36\textwidth]{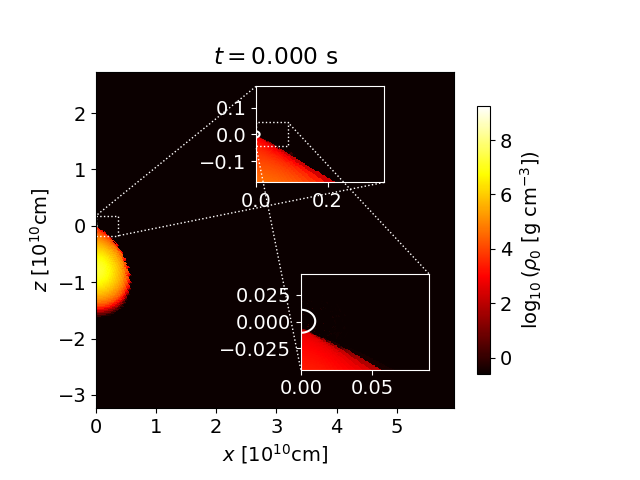}
    
    \includegraphics[width=0.36\textwidth]{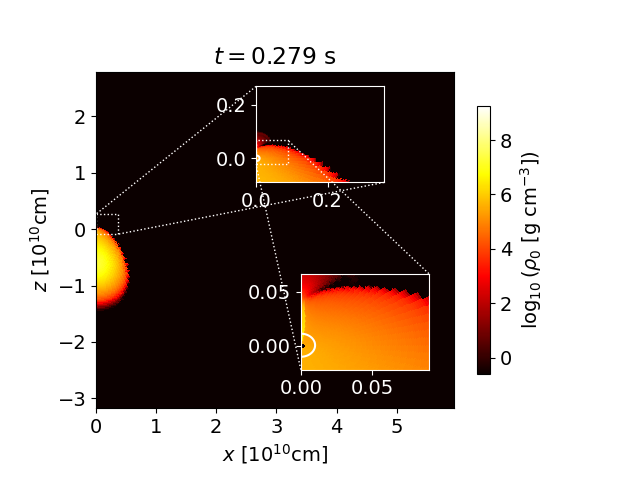}
    \includegraphics[width=0.36\textwidth]{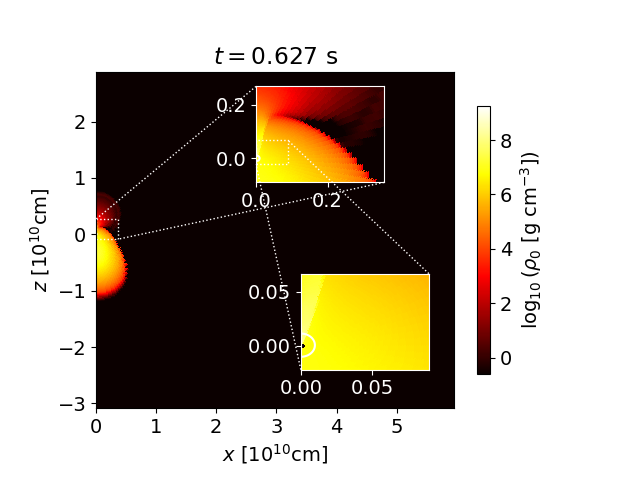}
    
    \includegraphics[width=0.36\textwidth]{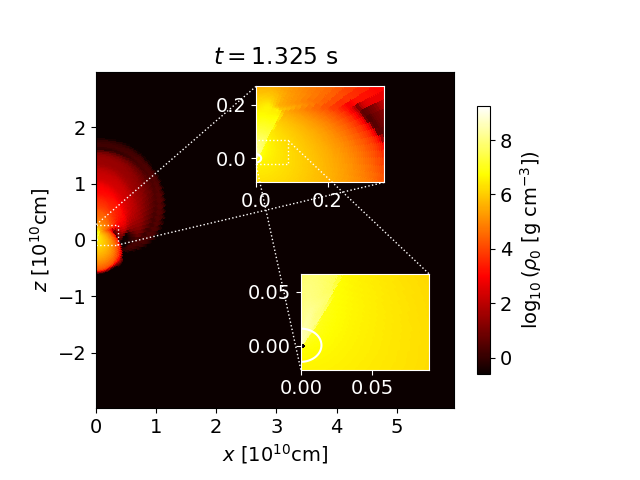}
    \includegraphics[width=0.36\textwidth]{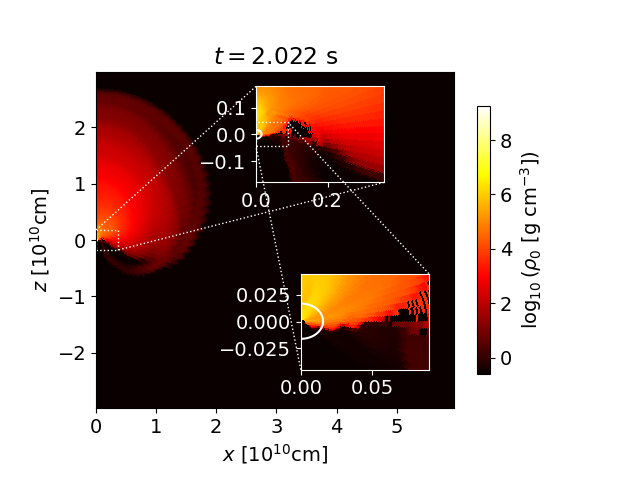}

    \includegraphics[width=0.36\textwidth]{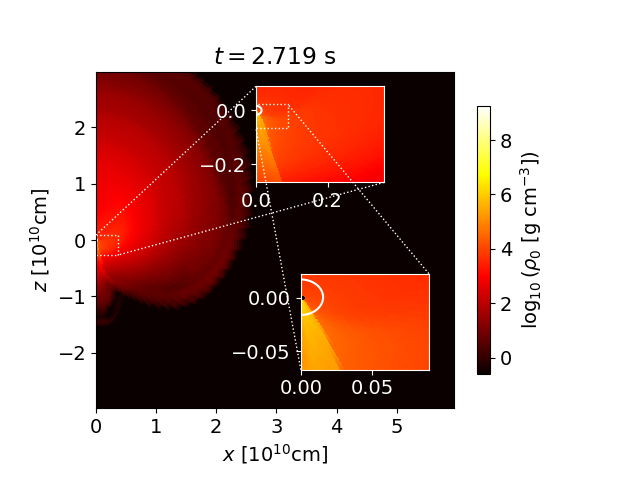}
    \includegraphics[width=0.36\textwidth]{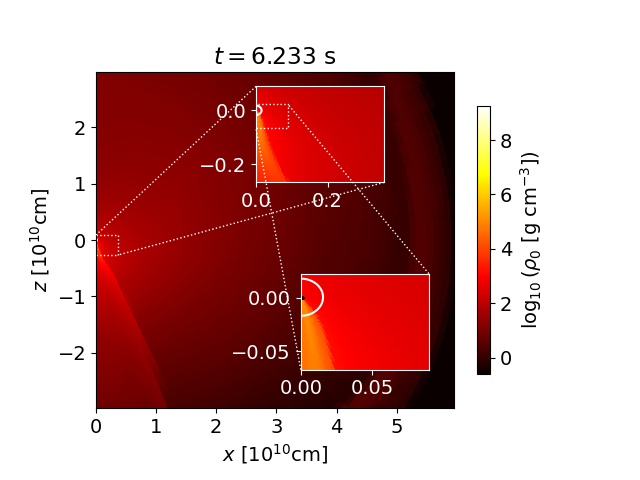}
    \caption{Snapshots of the rest-mass density $\rho_0$ for simulation A3 (see Table \ref{tab:indata}), i.e.~for the head-on collision of a black hole of initial mass $m(0) = 860 M_\odot$ and star of rest mass $430 M_{\odot}$.  The inserts zoom into the region close to the black hole.  See text for details and a discussion; see also \cite{animation} for animations.}
    \label{fig:density}
\end{figure*}

\begin{figure*}
    \centering
    \includegraphics[width=0.36\textwidth]{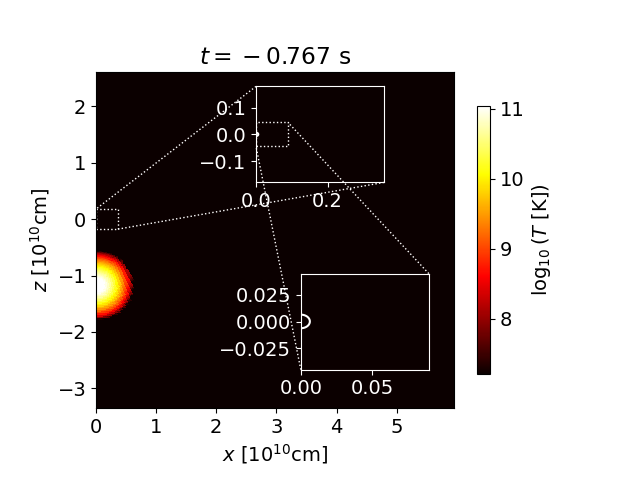}
    \includegraphics[width=0.36\textwidth]{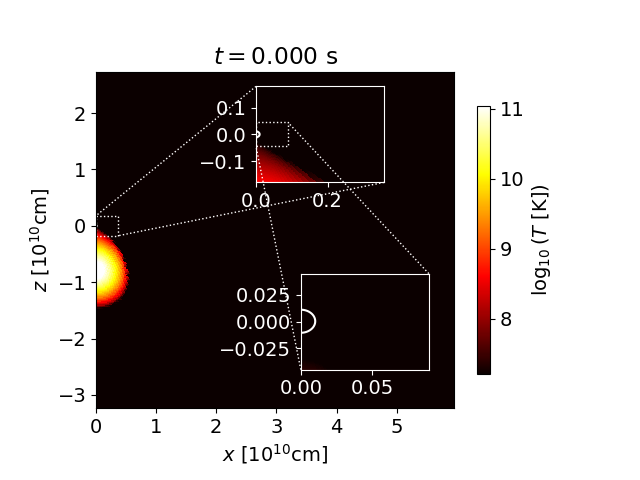}
    
    \includegraphics[width=0.36\textwidth]{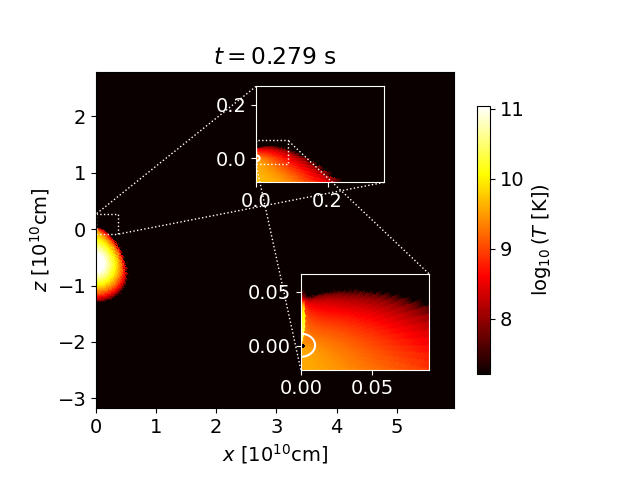}
    \includegraphics[width=0.36\textwidth]{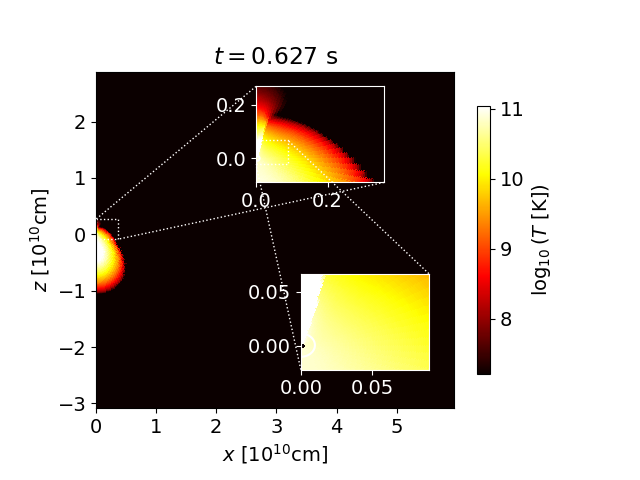}
    
    \includegraphics[width=0.36\textwidth]{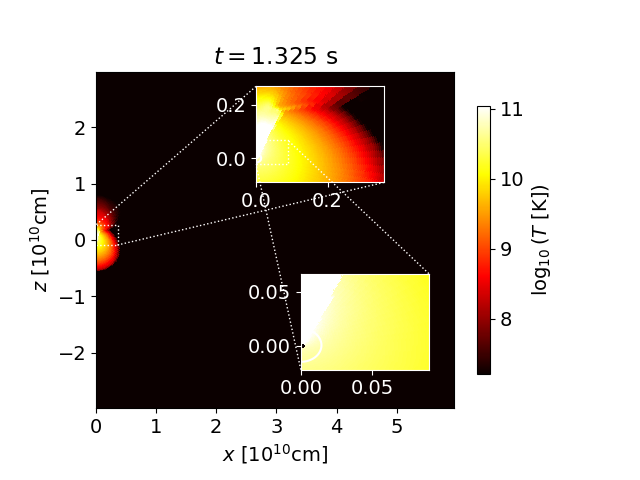}
    \includegraphics[width=0.36\textwidth]{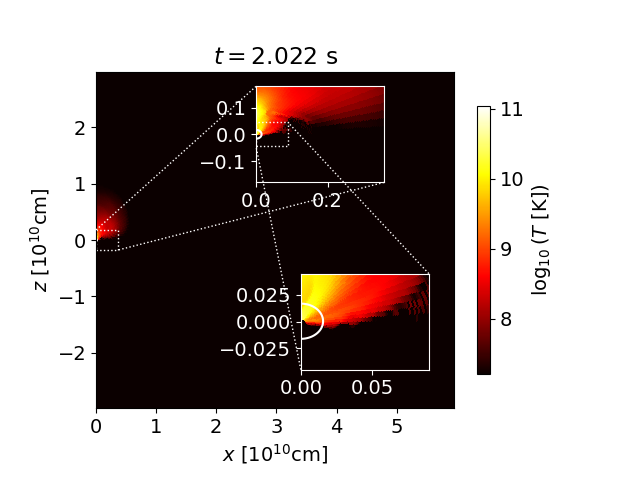}

    \includegraphics[width=0.36\textwidth]{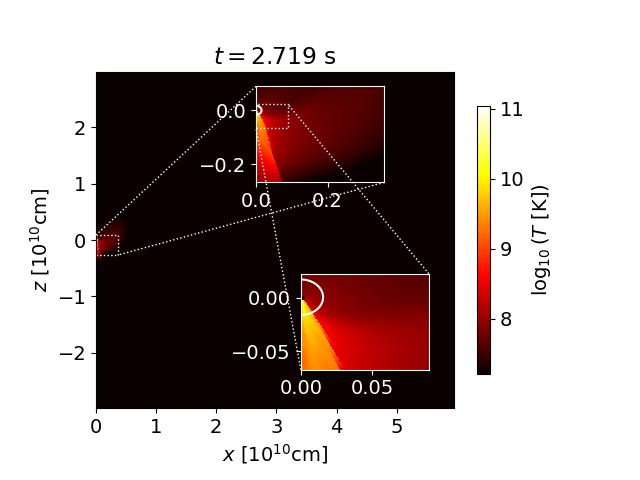}
    \includegraphics[width=0.36\textwidth]{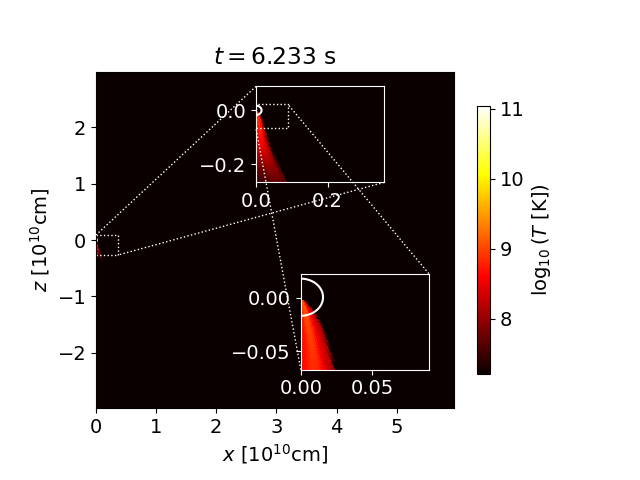}   
    \caption{Same as Fig.~\ref{fig:density}, but for the temperature $T$.}
    \label{fig:temperature}
\end{figure*}

We perform several simulations modeling the collision of the two stellar models A and B (see Table \ref{tab:star}) with black holes of different masses.  In Table \ref{tab:indata} we summarize our initial data configurations together with the final black hole masses as well as the rest mass remaining outside the black hole at the end of our simulations. Animations of some of these simulations can also be found at \cite{animation}.

As a representative example we show in Figs.~\ref{fig:density} and \ref{fig:temperature} snapshots of the density $\rho_0$ and the temperature $T$ for simulation A3, i.e.~the head-on collision of a black hole of mass $860 M_\odot$ with a star of mass of $430 M_\odot$.  The top-left panels in both figures show the initial data.  In the following we will choose the time $t$ so that the black hole first makes contact with the star at $t = 0$; therefore the initial data correspond to a negative time.  We also recall that our spherical polar coordinate system is dragged along by the black hole, so that the origin is always centered on the black hole.  We adjust the vertical $z$-axis in the eight main panels of Figs.~\ref{fig:density} and \ref{fig:temperature} so that they are centered on the approximate center of mass.  The inserts in each panel show the region around the black hole, whose apparent horizon is indicated by the white lines.  The innermost few gridpoints around the center of the black hole, where we excise the evolution of the matter variables, appear as black in the figures.

As the black hole and star approach each other the star deforms in the tidal field of the black hole.  This can be seen in the top right panels of Figs.~\ref{fig:density} and \ref{fig:temperature}, showing the moment when the two objects first come into contact.  Once the black hole penetrates the star, it leaves in its wake a shock-heated and over-dense region that is clearly visible in the second row of Figs.~\ref{fig:density} and \ref{fig:temperature}.   The pressure created by the high density and temperature launches a plume, resulting in the ejection of stellar material from the stellar surface. After approximately a stellar dynamical timescale (see the third row in the figures) most of the stellar material has been accreted by the black hole, while the ejected material forms an increasingly large cloud.  At later times the material ejected in the original plume appears to have wrapped around the black hole (bottom row), so that the black hole's wake now appears below the black hole and leads to the ejection of a secondary plume in the direction opposite to the original one.

\begin{figure}
    \centering
    \includegraphics[width=0.45\textwidth]{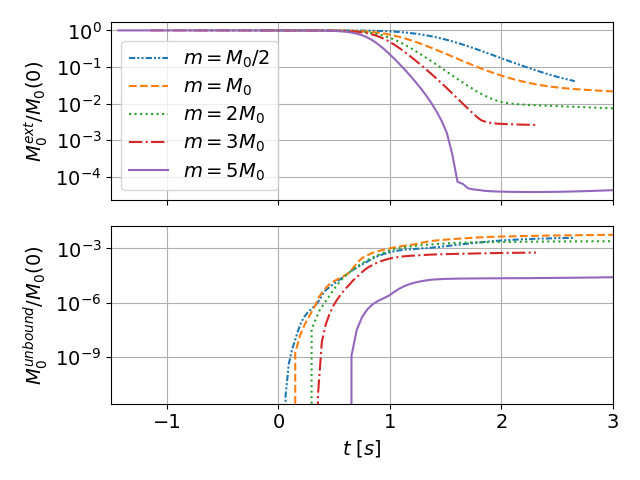}
    \caption{The rest mass $M_0^{\rm ext}$ outside of the black hole (top panel, see Eq.~[\ref{M_ext}]) and the unbound rest mass $M_0^{\rm unbound}$ (bottom panel), both as a fraction of the initial stellar rest mass $M_0(0)$, as a function of time $t$.  The time $t=0$ corresponds to the moment at which the black hole first comes into contact with the star. We show results for collisions with stellar model A, for which $M_0(0) = 430 M_\odot$ (see Table \ref{tab:indata}). The interor labels refer to the initial masses.}
    \label{fig:stellar_mass}
\end{figure}

\begin{figure}
    \centering
    \includegraphics[width=0.45\textwidth]{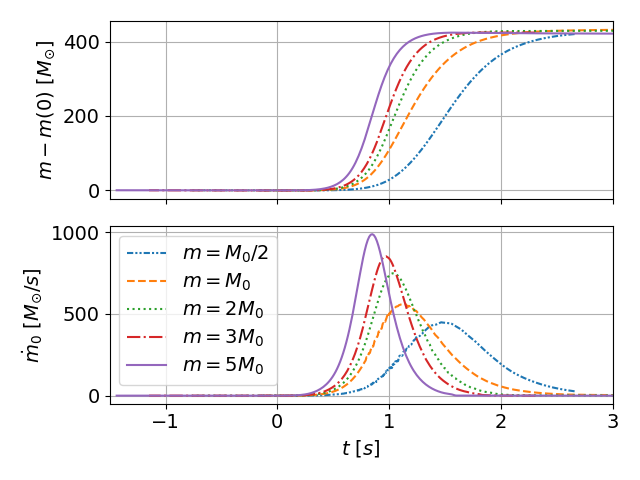}
    \caption{The black hole mass mass $m(t)$ (top panel, see Eq.~[\ref{m_irr}]) and the accretion rate $\dot m_0$ (bottom panel, see Eq.~[\ref{accretion_rate}]) as a function of time, for collisions with stellar model A.  In the top panel we show 
    the increase in the black hole's gravitational mass by subtracting its initial mass $m(0)$ from $m(t)$, so that all graphs start at zero and end at approximately $430 M_\odot$, the initial stellar mass.  The accretion rate shown in the bottom panel shows the rate of rest-mass accretion across the horizon, and therefore is only an approximate measure of the increase in the black hole's gravitational mass. The interor labels refer to the initial masses.}
    \label{fig:BH_mass}
\end{figure}

Simulations for different initial data configurations differ quantitatively, but they agree qualitatively in that the vast majority of the stellar material is accreted by the black hole, while only a small fraction of shock-heated material is ejected from the star.  In Figs.~\ref{fig:stellar_mass}  and \ref{fig:BH_mass} we demonstrate this for all collisions with star A.   Specifically, we show in the top panel of Fig.~\ref{fig:stellar_mass} the rest-mass $M_0^{\rm ext}$ remaining outside the black hole as a function of time, and in the bottom panel our estimate for the unbound rest-mass $M_0^{\rm unbound}$.  Fig.~\ref{fig:BH_mass} shows the increase of the black hole mass $m$ in the top panel, i.e.~the current black hole mass $m$ minus the initial black hole mass $m(0)$, and the rate of rest-mass accretion $\dot m_0$ in the bottom panel.  In all cases, at most a few percent of the initial stellar rest mass remain outside the black hole at the end of our simulations, and the black hole mass increases by values very similar to the initial stellar mass (i.e.~$M_0(0) = 430 M_\odot$ for star A).  

While our findings in Figs.~\ref{fig:stellar_mass} and \ref{fig:BH_mass} agree qualitatively, they do show quantitative differences.  In particular, in collisions with more massive black holes, more of the stellar mass is accreted, leaving less mass in the black hole exterior.  This is consistent with our estimates of Section \ref{sec:estimates}, which suggest that larger black holes are less likely to reemerge from the star, and accrete the stellar material more rapidly (see also Fig.~\ref{fig:estimate}, where larger black hole masses are further away from the critical lines, indicating that these trigger more immediate collapse than smaller black hole masses).  However, we caution that our results for small amounts of the mass ejected in the collision may well be affected by our atmosphere treatment in the stellar exterior (see Sect.~\ref{sec:dynamics}).   Smooth-particle treatments of hydrodynamics (see, e.g., \cite{RosD21,RosSND24} for fully relativistic examples) might provide an attractive alternative approach for tracking the detailed evolution and properties of the ejected plumes.

\begin{figure}
    \centering
    \includegraphics[width=0.45\textwidth]{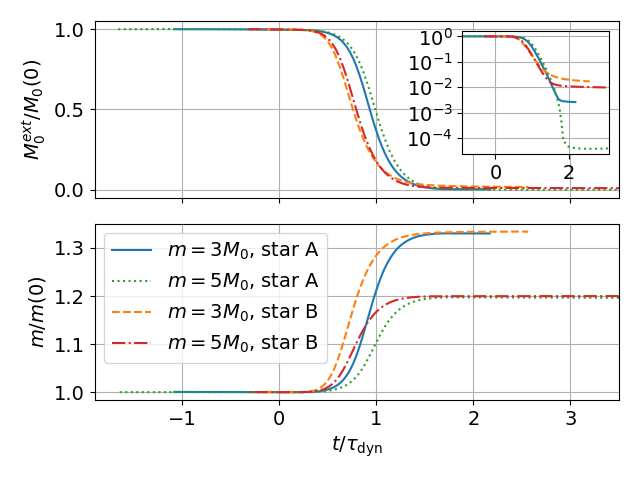}
    \caption{The rest mass $M_0^{\rm ext}$ remaining outside the black hole (top panel) and the black hole mass $m$ (bottom panel) as a function of time.  In all cases most of the mass is accreted by the black hole, whose mass increases by approximately by a factor $1 + M_0(0)/m(0)$, as expected.  We show the time $t$ in units of the dynamical timescale (\ref{tau_dyn}). The interior labels refer to the initial masses.}
    \label{fig:large_stars}
\end{figure}

In order to explore the effects of our choice of compaction of the initial stellar model we compare collisions with the two different stars A and B (see Table \ref{tab:star}) in Fig.~\ref{fig:large_stars}.  For each of the two stars we show results for collisions with two mass ratios, namely for black holes with initial masses of $m(0) = 3 M_0(0)$ and $m(0) = 5 M_0(0)$.  In the top panel of Fig.~\ref{fig:large_stars} we show the rest mass $M_0^{\rm ext}$ that remains in the black hole exterior as a function of time, and in the bottom panel the black hole mass $m$, both as a fraction of their initial masses.  At late times, $M_0^{\rm ext}/M_0(0)$ drops to small values, while $m/m(0)$ approaches $1 + M_0(0)/m(0)$, indicating that almost all of the stellar material is accreted and only a small fraction remains outside the black hole.  The exact value of $M_0^{\rm ext} / M_0$ does depend on the initial data, however, and, for both mass ratios, is larger for star B with the smaller compaction than star A (see insert in Fig.~\ref{fig:large_stars}).  We again caution that the details of the small amounts of ejecta may be affected by our numerical treatment and the artificial atmosphere.  Nevertheless, our findings do suggest that less material may be accreted, and hence more material returned to the interstellar medium, in collisions with less compact stars.  As we discussed in Section \ref{sec:init:OV}, simulating such stars in fully self-consistent general relativistic calculation poses a significant challenge.

\section{Summary and Discussion}
\label{sec:summary}

We report on fully relativistic simulations of the head-on collisions of IMBHs with VMSs.  As discussed by many authors, collisions and mergers of stars and black holes play an important role in the evolution of dense stellar clusters, which, in turn, may result in the formation of large seed black holes and their rapid growth to the SMBHs observed even in the early Universe.   In this paper we explore from first principles a key assumption made in many models of cluster evolution, namely that the head-on collision between a IMBH and a star leads to the entire star being accreted by the black hole.  

We model VMSs composed of hydrogen and helium, and take into account contributions to the pressure from both the Maxwell-Boltzmann gas and blackbody radiation.  We assume the stars to be isentropic initially, but choose models that are relatively compact in order to make our simulations feasible given our computational tools and resources.  We consider two different initial stellar models together with a range of initial black hole masses.  

While the details of the outcomes depend on both the mass ratio and the initial stellar compaction, they all agree qualitatively in that the vast majority of the stellar material does indeed end up inside the black hole.  In particular, we confirm analytic estimates that suggest that, for sufficiently large black holes, the black hole hole accretes almost the entire star during its first passage through the star.  

Our simulations also demonstrate that, once the black hole has penetrated the star, shock heating in its wake leads to an increase in density and temperature that results in the ejection of a small amount of gas from the stellar surface.  Initially this matter forms just a small plume ejected at the point of first contact, but, since some of this matter exceeds the escape speed and is hence unbound, it may return to the interstellar material in the stellar cluster. Our simulations are limited to head-on collisions, and we note again that tidal and possibly rotational effects could result in more mass being ejected from the star in more general parabolic or hyperbolic collisions.

\acknowledgments
It is a pleasure to thank Peter Diener for helpful discussions.  This work was supported in parts by National Science Foundation (NSF) grant PHY-2308821 to Bowdoin College, as well as NSF grants PHY-2006066 and PHY-2308242 to the University of Illinois at Urbana-Champaign.  Numerical simulations were performed on the Bowdoin Computational Grid.

\begin{appendix}

\section{Dynamical friction versus accretion drag}
\label{app:fric_and_drag}

In this brief appendix we provide a justification for assuming that hydrodynamical friction dominates over accretion drag in the supersonic regime.  To do so, we estimate the dynamical friction from (\ref{F_defl}) and the accretion drag from
\begin{equation}
F_{\rm acc} = \dot m v.
\end{equation}
For spherical accretion, the accretion rate $\dot m_{\rm sph}$ is given by the Bondi expression (\ref{Bondi}).  In order to account for the relative motion between the black hole and the gas we follow several previous authors \cite{Bon52,PetSST89,ShiMTS85} and multiply the spherical accretion rate with a Bondi-Hoyle-Lyttleton-like correction factor,
\begin{equation}
\dot m \simeq \dot m_{\rm sph} \left( \frac{a_s^2}{a_s^2 + v^2} \right)^{3/2}.
\end{equation}
We then compute the ratio between the dynamical friction and accretion drag to obtain
\begin{align}
\frac{F_{\rm defl}}{F_{\rm acc}} & 
\simeq \frac{4 \pi \rho G^2 m^2 \ln \Lambda}{v^2}
\left( \frac{4 \pi \rho G^2 m^2 \lambda}{a_s^3} \left(\frac{a_s^2}{a_s^2 + v^2} \right)^{3/2} v \right)^{-1} \nonumber \\
& = \frac{\ln \Lambda}{\lambda} \frac{a_s^3}{v^3}  \left(\frac{a_s^2 + v^2}{a_s^2} \right)^{3/2} = \frac{\ln \Lambda}{\lambda} \left(\frac{a_s^2 + v^2}{v^2} \right)^{3/2} \nonumber \\
& > \frac{\ln \Lambda}{\lambda},
\end{align}
where we have used $v > a_s$ in the last estimate.  Observing that $\ln \Lambda$ is expected to be on the order of 10, while $\lambda$ is of order unity (see Table~14.1 in \cite{ShaT83}), justifies our conclusion that dynamical friction dominates over accretion drag in the supersonic regime.

\section{Speed and Mach number of a black hole colliding head-on with a polytropic star}
\label{app:mach}

In this appendix we compute the speed and Mach number of a black hole intruder inside a Newtonian polytrope.  We assume that frictional forces can be ignored, so that the black hole's kinetic energy and linear momentum change only in response to the gravitational interaction with the star.  For simplicity we will also assume that the black hole mass is much smaller than the stellar mass, which is consistent with neglecting the frictional forces [see, e.g., Eq.~(\ref{F_defl})].

\begin{table}[]
    \centering
    \begin{tabular}{c|c|c|c|c|c}
         $n$ & $\Gamma$ & $\xi_n$ & $|\theta'(\xi_n)|$ & $\rho_c/\bar{\rho}$ & ${\mathcal M}_c$ \\
         \hline
         0.1 & 11.0 & 2.505 & 0.736 & 1.135 & 0.754 \\
         0.5 & 3.0 &       2.753 &  0.500 & 1.835 &  1.541\\
         1.0 & 2.0 & $\pi$ & $1/\pi$ & $\pi^2 / 3$ & 2.0 \\
         1.5 & 5/3 & 3.654 & 0.203 & 5.991 & 2.287\\
         2.0 & 3/2 &     4.353 & 0.127 & 11.403 & 2.493\\
         2.5 & 7/5 &     5.355 & 0.0768 & 23.407 & 2.656\\
         3.0 & 4/3 & 6.897 & 0.0424 & 54.183 & 2.785
    \end{tabular}
    \caption{Values of the Lane-Emden variables $\xi_n$ and $|\theta'(\xi_n)|$, the central concentration $\delta = \rho_c / \bar \rho$, and the central Mach number $\mathcal{M}_c$ for selected values of the polytropic index~$n$ (compare Table 4 in \cite{Cha39}).   }
    \label{tab:lane_emden}
\end{table}

For a Newtonian polytrope satisfying the polytropic EOS
\begin{equation}
P = K \rho^\Gamma = K \rho^{1 + 1/n}
\end{equation}
we may express the radius $r$ and the density $\rho$ in terms of the Lane-Emden variables $\xi$ and $\theta$,
\begin{align}
r & = a \xi, \\
\rho(r) &  = \rho_c \theta^n.
\end{align}
In the above $n = 1/(\Gamma - 1)$ is the polytropic index, $K$ the polytropic constant, $\rho_c$ the central density, and the constant $a$ is a lengthscale given by
\begin{equation}
a = \left[ \frac{(n+1) K \rho_c^{(1/n) - 1}}{4 \pi G} \right]^{1/2}.
\end{equation}
The variable $\theta$ then satisfies the Lane-Emden equation
\begin{equation} \label{lane_emden}
\frac{1}{\xi^2} \frac{d}{d\xi}\left( \xi^2 \frac{d \theta}{d \xi} \right) = - \theta^n
\end{equation}
(see, e.g., Section 3.3 in \cite{ShaT83} for a textbook treatment).  The stellar surface $r = R$, corresponding to $\theta = 0$, occurs at a location $\xi_n$ that depends on the polytropic index $n$.  Values for $\xi_n$, together with the derivative of $\theta$ at the surface and the central concentration $\delta = \rho_c / \bar \rho$, where $\bar \rho$ is the average density, ${\bar\rho} = M/(4 \pi R^3/3)$, have been tabulated by a number of different authors (see, e.g., \cite{Cha39}).  We reproduce a few selected values in Table \ref{tab:lane_emden}.

Using the surface values $\xi_n$ and $|\theta'(\xi_n)|$ as well as the Lane-Emden equation (\ref{lane_emden}) we may express the radius $r$ and the enclosed mass $m(r)$ as
\begin{equation} \label{r_and_m}
r = \frac{\xi}{\xi_n} \, R,~~~~~~~~~
m(\xi) = \frac{\xi^2 \theta'(\xi)}{\xi_n^2 \theta'(\xi_n)} \, M,
\end{equation}
where $M = m(R) = m(\xi_n)$ is the total mass.  We next compute the gravitational potential $\phi$ in the stellar interior, normalized to zero at infinity, from
\begin{align} 
\phi(r) & = - \frac{M}{R} + \int_R^r \frac{m(r)}{r^2} dr
= - \frac{M}{R} + \frac{M}{R} \int_{\xi_n}^\xi \frac{\theta'(\xi)}{\theta'(\xi_n)} \frac{d\xi}{\xi_n} \nonumber 
\\
& = - \frac{M}{R} \left( 1 - \frac{1}{\xi_n \theta'(\xi_n)} \int_{\xi_n}^\xi d \theta'(\xi) \right) \nonumber \\
& = - \frac{M}{R} \left( 1 + \frac{\theta(\xi)}{\xi_n |\theta'(\xi_n)|} \right),  \label{potential1}
\end{align}
where we have used (\ref{r_and_m}) in the second equality and $\theta(\xi_n) = 0$ together with $\theta'(\xi_n) < 0$ in the last.  Finally, we note that
\begin{equation}
\frac{M}{R} = \frac{(n + 1) K}{G} \rho_c^{1/3} \xi_n |\theta'(\xi_n)|
\end{equation}
(see, e.g., Eqs.~(3.3.9) and (3.3.10) in \cite{ShaT83}) to express (\ref{potential1}) as
\begin{equation} \label{potential}
\phi = - \frac{(n+1)K}{G} \rho_c^{1/n} \Big( \xi_n |\theta'(\xi_n)| + \theta(\xi) \Big).
\end{equation}
For an object in free-fall, starting from rest at infinity, energy conservation then yields the speed
\begin{equation} \label{v}
v = (- 2 \phi)^{1/2}.
\end{equation}

We next write the (Newtonian) sound speed $c_s$ as
\begin{align}
    c_s & = \left( \frac{dP}{d\rho} \right)^{1/2} 
    = \left( \frac{n + 1}{n} K \rho^{1/n} \right)^{1/2} \nonumber \\
    & = \left( \frac{n + 1}{n} K \rho_c^{1/n} \theta(\xi) \right)^{1/2}.  \label{sound}
\end{align}
Combining (\ref{sound}) with the speed (\ref{v}), using (\ref{potential}), we write the Mach number ${\mathcal M} \equiv v / c_s$ as
\begin{equation} \label{mach1}
\mathcal{M} = \left[ 2 n \left(1 + \frac{\xi_n | \theta'(\xi_n)|}{\theta(\xi)} \right) \right]^{1/2}.
\end{equation}
At the surface, where $\theta(\xi_n) = 0$, the Mach number is infinite.  At the stellar center, where the sound speed takes a maximum, the Mach number takes a minimum.  We evaluate this term using $\theta(0) = 1$, for which (\ref{mach1}) reduces to
\begin{equation} \label{mach2}
\mathcal{M}_c = \Big( 2 n \left(1 + \xi_n | \theta'(\xi_n)| \right) \Big)^{1/2}.
\end{equation}
For an $n = 1$ ($\Gamma = 2$) polytrope  we have $\xi_1 | \theta'(\xi_1)| = 1$ and hence $\mathcal{M}_c = 2$, which is what we had previously derived in Appendix C.1 of \cite{BauS24c}.  For an $n = 3$ ($\Gamma = 4/3$) polytope, which serves as a good model for a VMS, we have $\xi_3 | \theta'(\xi_3) | = 0.2926$ and hence
\begin{equation}
\mathcal{M}_c = 2.785~~~~~~~~~~~~~~~~~~~(n = 3),
\end{equation}
indicating that the black hole will remain supersonic throughout the star.  For the large black hole masses considered in this paper, however, retarding frictional forces cannot be ignored, as demonstrated by both the estimates in Section \ref{sec:estimates} and the numerical results in Section \ref{sec:results}.

\section{Numerical evaluation of the EOS}
\label{app:EOS_num}

In this appendix we provide some details regarding the entropy of the hydrogen and helium gas, suitable units for our numerical simulation, and the calculation of the speed of sound, including the radiation.

\subsection{Entropy}
\label{sec:EOS_num:entropy}

The total gas entropy $S_g$ is the sum of the entropies of each free species, in our case the hydrogen and helium atoms as well as the electrons,
\begin{equation} \label{app_total_entropy}
S_g = S_{\rm H} + S_{\rm He} + S_{\rm e}.
\end{equation}
For each specie $i$ we have
\begin{equation} \label{app_entropy_specie}
S_i = N_i k_B \left[ \ln \left( \frac{g_i V}{N_i} \left( \frac{4 \pi m_i U_i}{3 N_i h^2} \right)^{3/2} \right) + \frac{5}{2} \right],  
\end{equation}
where $N_i$ is the total number of specie $i$ particles, $g_i$ their degeneracy, $V$ the volume of the gas, $m_i$ the mass of the specie particle, and $U_i = 3 N_i k_B T / 2$ the internal energy of specie $i$.  In the following we use $m_{\rm H} = m_B$ and $m_{\rm He} = 4 m_B$, where $m_B$ is the baryon mass, as well as $g_{\rm H} = g_e = 2$ and $g_{\rm He} = 1$.

The gas entropy per baryon $s_g$ is then given by $s_g = S_g / N$, where $N$ is the total number of baryons.  Using
\begin{align}
\frac{N_{\rm H}}{N} = X,~~~
\frac{N_{\rm He}}{N} = \frac{Y}{4},~~~
\frac{N_e}{N} = X + \frac{Y}{2}
\end{align}
for a fully ionized gas, as well as $\rho_0 = m_B n = m_B N / V$ we find, after some algebra, 
\begin{equation} 
s_g = \frac{k_B}{\mu} \ln \frac{(k_B T)^{3/2}}{\rho_0} + s_0
\end{equation}
where the mean molecular weight $\mu$ is given by (\ref{mu}) and the constant $s_0$ by
\begin{align} \label{app_entropy_constant}
\frac{s_0}{k_B} = & \frac{3}{2 \mu} \ln \frac{2 \pi}{h^2} + \frac{5}{2 \mu}
+ \left( 2 X + \frac{7Y}{4} \right) \ln 2
\\
& - \left( X + \frac{Y}{2} \right) \ln \left(X + \frac{Y}{2} \right) 
- X \ln X - \frac{Y}{4} \ln Y \nonumber \\ 
& + \left( \frac{3X}{2} + \frac{3 Y}{4} \right) \ln m_e
+ \left( \frac{7 X}{2} + \frac{9 Y}{8} \right) \ln m_B. \nonumber 
\end{align}
For a pure hydrogen gas with $X = 1$ and $Y = 0$ the above expressions reduce to Eq.~(17.3.4) in \cite{ShaT83}.

\subsection{Units}
\label{sec:EOS_num:units}

As in Section \ref{sec:diagnostics} we adopt  geometrized units with $G = 1 = c$.  We then
introduce dimensionless variables by first writing the entropy in units of the Boltzmann constant $k_B$,
\begin{equation}
\bar s \equiv \frac{s}{k_B},
\end{equation}
and the temperature $T$ in units of $m_e / k_B = 5.93 \times 10^9$~K,
\begin{equation}
\bar T \equiv \frac{k_B T}{m_e}.
\end{equation}
Further defining a rescaled radiation constant
\begin{equation}
\alpha \equiv \frac{m_e^4}{k_B^4} a = \frac{8 \pi^5 m_e^4}{15 h^3}
= 7.73 \times 10^{-25} \mbox{cm}^{-2}
\end{equation}
we may write the pressure (\ref{pressure_1}) as
\begin{equation}
P_r = \frac{1}{3} \alpha \bar T^4 + 
\frac{m_e}{m_B} \frac{\rho_0 \bar T}{\mu},
\end{equation}
and the entropy (\ref{entropy_1}) as
\begin{equation} \label{entropy_2}
\bar s = \frac{m_B}{m_e} \frac{4 \alpha \bar T^3}{3 \rho_0}
+ \frac{1}{\mu} \ln \left( 
\frac{\bar T^{3/2}}{\rho_0} \right) + \bar s_0 + \frac{3}{2 \mu} \ln m_e.
\end{equation}

Having expressed temperature and entropy in terms of non-dimensional variables already, we now choose
\begin{equation} \label{length_scale}
L \equiv 10^6 M_\odot = 1.49 \times 10^{11} \mbox{cm} = 4.99 \, \mbox{s}
\end{equation}
as a typical length scale and then define new dimensionless quantities according to 
\begin{equation}
\begin{array}{rclrcl}
\bar P & \equiv & L^2 \, P, ~~~~~~~~~~~
& \bar \rho & \equiv & L^2 \, \rho, \\
\bar m_B & \equiv & L^{-1} \, m_B, 
& \bar \alpha & \equiv & L^2 \, \alpha,
\end{array}
\end{equation}
and similar for other terms. We note that the specific internal energy density $\epsilon$ is dimensionless already in geometrized units, and therefore does not get rescaled with $L$.  In terms of these, the pressure, specific internal energy density, total mass-energy density, and entropy become
\begin{subequations} \label{nd}
\begin{align}
\bar P & = \frac{1}{3} \bar \alpha \bar T^4 + \frac{m_e}{m_B} \frac{\bar \rho_0 \bar T}{\mu}, \label{pressure_nd} \\
\epsilon & = \frac{\bar \alpha \bar T^4}{\bar \rho_0} +  \frac{3}{2} \frac{m_e}{m_B} \frac{\bar T}{\mu}, \label{epsilon_nd} \\
\bar \rho & = \bar \rho_0 + \bar \alpha \bar T^4 + \frac{3}{2} \frac{m_e}{m_B} \frac{\bar \rho_0 \bar T}{\mu}, \label{rho_nd}
\end {align}
and
\begin{align}
\bar s & = \frac{m_B}{m_e} \frac{4 \bar \alpha \bar T^3}{3 \bar \rho_0} + \frac{1}{\mu} \ln \left( \frac{\bar T^{3/2}}{\bar \rho_0} \right) + \bar s_e. \label{s_nd}
\end{align}
\end{subequations}
Here the dimensionless constant $\bar s_e$ is given by 
\begin{align}
\bar s_e & \equiv \bar s_0 + \frac{3}{2 \mu} \ln m_e + \frac{2}{\mu} \ln L \\
& = \frac{3}{2 \mu} \ln \frac{2 \pi}{\bar h^2} + \frac{5}{2 \mu}
+ \left( 2 X + \frac{7Y}{4} \right) \ln 2
\nonumber \\
& - \left( X + \frac{Y}{2} \right) \ln \left(X + \frac{Y}{2} \right) 
- X \ln X - \frac{Y}{4} \ln Y \nonumber \\ 
& + \left( \frac{9X}{2} + \frac{15 Y}{8} \right) \ln \bar m_e
+ \left( \frac{7 X}{2} + \frac{9 Y}{8} \right) \ln \bar m_B, \nonumber 
\end{align}
where the nondimensional Planck constant $\bar h \equiv L^{-2} h$ should not be confused with $\hbar$. 

\subsection{Sound speed}
\label{sec:EOS_num:sound}

The sound speed $c_s$ is given by
\begin{equation} \label{cs_1}
c_s^2 = \left( \frac{dP}{d\rho} \right)_s
= \left( \frac{d \bar P}{d \bar \rho} \right)_{\bar s}
= \left( \frac{d \bar P}{d \bar \rho_0} \right)_{\bar s}
\left( \frac{d \bar \rho_0}{d \bar \rho} \right)_{\bar s}
\end{equation}
where the subscript $s$ outside the parantheses indicates that changes are to be taken at constant entropy $s$.  From the first law of thermodynamics for isentropic processes with $ds = 0$,
\begin{equation}
d \epsilon = - P d \left( \frac{1}{\rho_0} \right) 
= \frac{\bar P}{\bar \rho_0^2} d \bar \rho_0,
\end{equation}
we obtain
\begin{equation}
d \bar \rho = d\left( \bar \rho_0 (1 + \epsilon)\right) = H d \bar \rho_0
\end{equation}
and hence
\begin{equation}
\left( \frac{d \bar \rho_0}{d \bar \rho} \right)_{\bar s} = \frac{1}{H},
\end{equation}
where $H$ is the enthalpy
\begin{equation}
H = 1 + \epsilon + \frac{\bar P}{\bar \rho_0}
\end{equation}
(we use an upper case letter $H$ rather than the more customary lower case $h$ in order to distinguish from Planck's constant).  We may therefore write the sound speed (\ref{cs_1}), including the radiation, as
\begin{equation} \label{cs_2}
c_s^2 = \frac{1}{H} \left( 
\left( \frac{d \bar P_r}{d \bar \rho_0} \right)_{\bar s} + 
\left( \frac{d \bar P_g}{d \bar \rho_0} \right)_{\bar s} 
\right).
\end{equation}

In order to evaluate the two terms on the right-hand side of (\ref{cs_2}) we first note that 
\begin{equation} \label{ds_r}
d\bar s_r = d \left( \frac{4 \bar \alpha \bar T^3}{3 \bar \rho_0} \right) 
= 3 \bar s_r \frac{d \bar T}{\bar T} 
- \bar s_r \frac{d \bar \rho_0}{\bar \rho_0}
\end{equation}
and 
\begin{equation} \label{ds_g}
d\bar s_g = d \left( \ln \left( \bar \kappa \frac{\bar T^3}{\bar \rho_0^2} \right) + 5  \right)
= 3 \frac{d \bar T}{\bar T} - 2 \frac{d \bar \rho_0}{\bar \rho_0}.
\end{equation}
For isentropic changes we must have $d \bar s_r = - d \bar s_g$, from which we obtain
\begin{equation}
\frac{d \bar T}{\bar T}(3 \bar s_r + 3) 
= \frac{d \bar \rho_0}{\bar \rho_0} (\bar s_r + 2)
\end{equation}
or
\begin{equation} \label{dTdrho0}
\left( \frac{d \bar T}{d \bar \rho_0} \right)_{\bar s}
= \sigma_s \frac{\bar T}{\bar \rho_0},
\end{equation}
where we have abbreviated
\begin{equation} \label{sigma}
\sigma_s \equiv \frac{\bar s_r + 2}{3 \bar s_r + 3}.
\end{equation}

The first term on the right-hand side of (\ref{cs_2}) can now be written as
\begin{equation}
\left( \frac{d \bar P_r}{d \bar \rho_0} \right)_{\bar s}
= \frac{\partial \bar P_r}{\partial \bar \rho_0} 
+ \frac{\partial \bar P_r}{\partial \bar T}
\left( \frac{d \bar T}{d \bar \rho_0} \right)_{\bar s}.
\end{equation}
Here the first term on the right-hand side vanishes, and we evaluate the second from $\bar P_r = \bar \alpha \bar T^4/3$ together with (\ref{dTdrho0}) to obtain
\begin{equation}
\left( \frac{d \bar P_r}{d \bar \rho_0} \right)_{\bar s}
= \frac{4}{3} \sigma_s \frac{\bar \alpha \bar T^4}{\rho_0} 
= \frac{4}{3} \sigma_s \epsilon_r.
\end{equation}
We similarly use $\bar P_g = 2 \bar \rho_0 \bar T$ to evaluate the second term on the right-hand side of (\ref{cs_2}),
\begin{align}
\left( \frac{d \bar P_g}{d \bar \rho_0} \right)_{\bar s}
& = \frac{\partial \bar P_g}{\partial \bar \rho_0} 
+ \frac{\partial \bar P_g}{\partial \bar T}
\left( \frac{d \bar T}{d \bar \rho_0} \right)_{\bar s} \nonumber
= 2 \bar T + 2 \bar \rho_0 \sigma_s \frac{\bar T}{\bar \rho_0} \\
& = 2 (1 + \sigma_s) \bar T = \frac{2}{3} (1 + \sigma_s) \, \epsilon_g. 
\end{align}
Collecting terms we may now write the sound speed as
\begin{equation} \label{cs}
c_s^2 = \frac{2}{3 H} \left( 2\sigma_s \epsilon_r 
+ (1 + \sigma_s)\, \epsilon_g \right).
\end{equation}

\end{appendix}


%

\end{document}